\documentstyle[12pt,aaspp4]{article}

\def\simlt{\ {\raise-.5ex\hbox{$\buildrel<\over\sim$}}\ }
\def\simgt{\ {\raise-.5ex\hbox{$\buildrel>\over\sim$}}\ }

\def\asca{{\it ASCA\/}}

\def\axaf{{\it AXAF\/}}

\def\einstein{{\it Einstein\/}}

\def\heao1{{\it HEAO-1\/}}

\def\rosat{{\it ROSAT\/}}
\def\rxte{{\it RXTE\/}}

\def\xmm{{\it XMM\/}}
\def\vela5b{{\it Vela-5B\/}}

\begin{document} 


\title{Exploratory ASCA Observations of Broad Absorption Line Quasi-Stellar Objects}

\author{S. C. Gallagher, W. N. Brandt and R. M. Sambruna}
\affil{Department of Astronomy and Astrophysics, Penn State University,
525 Davey Lab, University Park, PA 16802}

\author{S. Mathur}
\affil{Harvard-Smithsonian Center for Astrophysics, 60 Garden Street, 
Cambridge, MA 02138}
 
\author {N. Yamasaki}
\affil{Department of Physics, Faculty of Science, 
Tokyo Metropolitan University, 1-1 Minami-Osawa, 
Hachioji, Tokyo 192-03, Japan}

\begin{abstract}
We present the analysis and interpretation of a sample of eight \asca\ 
observations of Broad Absorption Line Quasi-Stellar Objects (BALQSOs). 
This is the first moderate-sized sample of sensitive BALQSO observations 
above 2~keV, and the BALQSOs in our sample are among the optically 
brightest known ($B=$~14.5--18.5). Despite the ability of 
2--10~keV X-rays to penetrate large column densities, we find BALQSOs 
to be extremely weak sources above 2~keV, and we are only able to add
two new 2--10~keV detections (0226--104 and IRAS~07598+6508) to 
those previously reported. 
By comparison with non-BALQSOs of similar optical continuum
magnitudes, we derive the column densities needed to suppress
the expected X-ray fluxes of our BALQSOs. In several cases we derive 
column densities $\simgt 5\times 10^{23}$~cm$^{-2}$ for a neutral absorber
with solar abundances. These are the largest X-ray column densities
yet inferred for BALQSOs, and they exceed \rosat\ lower limits by 
about an order of magnitude. 

Optical brightness does not appear to be a good predictor of 
2--10~keV brightness for BALQSOs, but our data do suggest that the 
BALQSOs with high optical continuum polarizations {\it may\/} be the
X-ray brighter members of the class. For example, the highly 
polarized object PHL~5200 appears to be unusually X-ray bright for a 
BALQSO given its optical magnitude. 

We discuss the implications of our results for future observations
with \axaf\ and \xmm. If the objects in our sample are representative
of the BALQSO population, precision X-ray spectroscopy of most BALQSOs
will unfortunately prove difficult in the near future. 
\end{abstract}



\section{Introduction}

While Broad Absorption Line Quasi-Stellar Objects (BALQSOs) allow us
to view substantial and energetically important gas outflows that are
probably present in most QSOs (e.g. Weymann 1997), 
their study in the X-ray regime has
not yet matured. BALQSOs are known to be much weaker in the soft 
X-ray band ($<2$~keV) than QSOs that lack BALs
(e.g. Kopko et~al. 1994; Green et~al. 1995; 
Green \& Mathur 1996, hereafter GM96), 
and only a few BALQSOs have reliable X-ray detections. Their low
soft-X-ray fluxes may arise as a result of 
photoelectric X-ray absorption by the same 
outflowing matter that makes the BALs in the ultraviolet.
Hard X-ray ($>2$~keV) observations of BALQSOs can in principle test 
the absorption hypothesis as well as constrain the properties, in 
particular the column densities, of BALQSO outflows. Column density 
constraints for BALQSOs are difficult to obtain from ultraviolet data
due to BAL saturation (e.g. Arav 1997; Hamann 1998), and the 
limited X-ray data now available suggest that previous column
density estimates from ultraviolet lines were too small by a factor of
$\approx 100$ or more. Because the photoelectric X-ray absorption 
cross section is a strongly decreasing function of energy, 
BALQSOs that are weak in the soft X-ray band (e.g. for \rosat) could 
be much brighter at higher X-ray 
energies (e.g. for \asca).\footnote{The \rosat\ band is from 0.1--2.4~keV 
and peaks from 0.9--1.4~keV, while the \asca\ band is from 
0.6--9.5~keV and has high sensitivity in the 3--7~keV range. 
To good approximation, the photoelectric X-ray absorption cross section 
is proportional to $E^{-{8\over 3}}$. Thus a 6~keV X-ray is $\approx 120$ times 
more penetrating than a 1~keV X-ray.}
In the hard X-ray band, only the BALQSOs PHL~5200 ($z=1.98$, $B\approx 18.5$)
and Mrk~231 ($z=0.042$, $B\approx 14.5$) have been studied in detail. 
Mathur, Elvis \& Singh (1995) argue for the presence of heavy X-ray 
absorption with $N_{\rm H}\sim 1\times 10^{23}$~cm$^{-2}$ in 
PHL~5200, although the observed flux is low making reliable
spectral analysis difficult (see \S3 for further discussion). 
The \asca\ data for Mrk~231 also suggest a large intrinsic column density 
($\sim 2\times 10^{22}$--$10^{23}$~cm$^{-2}$; Iwasawa 1999; Turner 1999).

Since at least $\approx 10$\% and perhaps up to 30--50\% of all
QSOs have BAL gas along our line of sight (e.g. Goodrich 1997; 
Krolik \& Voit 1998), our lack of knowledge about the X-ray 
properties of BALQSOs represents a serious deficiency.
In an attempt to remedy this situation, we performed \asca\ observations 
of several of the BALQSOs that seemed most likely to be X-ray bright above 
2~keV. Given the unknown hard X-ray properties (e.g. fluxes) of BALQSOs, we 
adopted a conservative strategy of making moderate-depth `exploratory'  
observations of a fairly large number of objects. In this manner, we
could learn about the 2--10~keV properties of as many BALQSOs as possible
without being too heavily invested in the uncertain results from any one 
object. BALQSOs found to be reasonably X-ray bright based on  
exploratory observations could then be followed up with deeper X-ray 
spectroscopic observations. We chose as our targets some 
of the optically brightest (in the $B$ and $R$ bands) BALQSOs,
since there is a general correlation between optical brightness and
intrinsic X-ray brightness for QSOs (e.g. Zamorani et~al. 1981). 
In this paper we present \asca\ results for five of the optically 
brightest BALQSOs known: 
PG 0043+039, 
0043+008,
0226--104, 
PG~1700+518 and 
LBQS~2111--4335.
In the $B$ band, these objects are all brighter than the 
\asca-detected BALQSO PHL~5200. 
In addition, we include the \asca\ results from 
archived observations of
IRAS~07598+6508, 
Mrk~231 and 
PHL~5200.
Redshifts, $B$ and $R$ magnitudes, and Galactic column densities for 
the BALQSOs in our sample are given in Table~1. Many of our BALQSOs are 
comparably bright in the optical band to radio-quiet (RQ) QSOs which 
\asca\ has studied in significant detail (e.g. PG~1634+706 and PG~1718+481, 
Nandra et~al. 1995; IRAS~13349+2438, Brandt et~al. 1997), and
our sample is on average significantly brighter at $B$ than the 
high-redshift RQ QSOs studied with \asca\ by 
Vignali et~al. (1999). Aside from being some of the optically
brightest BALQSOs in the sky, the objects in our sample span a range
of other properties (e.g. absorption line strength, absorption line
shape, redshift, optical continuum polarization, infrared luminosity). 
While they do not form a rigorously defined complete
sample, they do appear to comprise a reasonably representative 
subsample of the BALQSO population as a whole. 

As mentioned above, if BALQSOs suffer from heavy X-ray absorption
then they will be much brighter above 2~keV than at lower 
energies. For example, a hypothetical BALQSO at $z=2$ with an 
absorption column density of $5\times 10^{23}$~cm$^{-2}$ is expected 
to be $\approx 20$ times brighter in the 2--10~keV band than in the
0.1--2.0~keV band (observed-frame with a photon index of $\Gamma=2.0$ for 
the underlying power law). The 2--10~keV sensitivity of \asca\ thus 
makes it a superior tool to \rosat\ for studying BALQSOs with X-ray 
column densities in the wide range from
$\approx$~(1--500)$\times 10^{22}$~cm$^{-2}$ (column densities
substantially larger than $\sigma_{\rm T}^{-1}=1.50\times 10^{24}$~cm$^{-2}$
are optically thick to Thomson scattering and are thus impenetrable 
to X-rays with energies below $m_{\rm e}c^2$). 
Furthermore, even moderate-length \asca\ non-detections can set 
important constraints on the 2--10~keV fluxes and internal column 
densities of BALQSOs. These are important for planning 
\axaf/\xmm\ spectroscopic observations, and in some cases \asca\ 
non-detections can raise the current \rosat\ lower limits on 
BALQSO X-ray column densities by roughly an order of magnitude 
(from $\approx 5\times 10^{22}$~cm$^{-2}$ to $\approx 
5\times 10^{23}$~cm$^{-2}$). If the outflowing BAL material causes 
the inferred X-ray absorption, this increases the implied BALQSO 
mass outflow rates and kinetic luminosities.


\section{Observations and Data Analysis}

\subsection{Observation Details and Data Reduction}

In Table~1 we list the relevant observation dates, exposure times
and instrument modes for our objects. The data resulting from these 
observations were reduced and analyzed with {\sc ftools} and {\sc xselect}
following the general procedures described in Brandt et~al. (1997). We have 
used Revision 2 data (Pier 1997) and adopted the standard Revision~2 
screening criteria.

\subsection{Image Analyses and Count Rate Constraints}

We used {\sc xselect} to create
full (0.6--9.5~keV for SIS and 0.9--9.5~keV for GIS) 
and hard (2--9.5~keV) band images for each of the four \asca\ detectors. 
We also created summed SIS0+SIS1 and GIS2+GIS3 images in the
full and hard bands. Image analysis was performed using {\sc ximage}
(Giommi, Angelini \& White 1997). We first 
attempted to independently check the \asca\ astrometry  
using serendipitous X-ray sources within the fields of view. We 
correlated \asca\ serendipitous source positions with objects in 
coincident \rosat/\einstein\ images as well as the {\sc ned}/{\sc simbad} 
databases. We were able to independently verify the astrometry for
the observations of all BALQSOs but PG~0043+039, Mrk~231 and PHL~5200. We 
comment that the \asca\ astrometry is generally quite reliable 
(see Gotthelf 1996), and our independent checking is only done for 
confirmation. 

We then searched all the images described above for any
significant X-ray sources that were positionally coincident with
the precise optical positions of our BALQSOs.  
In order to determine the observed SIS count rate or upper limit for a 
given BALQSO, the 0.6--9.5~keV counts were extracted from a circular target region 
with a $3.2^{\prime}$ radius centered on the optical BALQSO position. This provided 
counts for the target region, $N_{\rm t}$. An annular or circular source-free 
background region was chosen and similarly extracted to give background 
counts. The background counts were normalized to the area 
of the target region to obtain $N_{\rm b}$ which was then subtracted from $N_{\rm t}$.  
If the difference was $<3\sigma$ where $\sigma=\sqrt{N_{\rm b}}$, 
the observation was considered a non-detection. In this case an upper limit on the 
count rate was taken to be $3\sqrt{N_{\rm t}}/t_{\rm e}$ where $t_{\rm e}$ is the 
exposure time listed in Table~1. For observations with $\geq 3\sigma$ detections, 
the count rate was calculated as $(N_{\rm t}-N_{\rm b})/t_{\rm e}$.  
A similar procedure was followed for the GIS detectors. However, for the 
GIS the extracted energy range was from 0.9--9.5~keV, and the radius of the 
target region was $5^{\prime}$. The results of these analyses are presented 
in Table~2, and we comment on special cases in \S3. Only four of 
these optically bright BALQSOs were detected with high statistical 
significance: 0226--104, IRAS~07598+6508, Mrk~231 and PHL~5200. 
We have chosen our target region sizes based upon \S7.4 of 
{\it The ASCA Data Reduction Guide\/}, and we have found that
varying the target region sizes within the recommended ranges does 
not materially change our basic results. 
When calculating $N_{\rm b}$ we have investigated at least two 
background regions for each SIS/GIS image, and we find that our 
results also do not materially depend upon our choice of background 
region. 
When our SIS observations were made using 2~CCD mode (see Table~1), 
we took all background photons from the chip that contained the
target. 

We also give observed frame 2--10~keV fluxes or upper limits for our
BALQSOs in Table~2. These were computed for SIS0 with 
{\sc pimms} (Mukai 1997) using a power law with $\Gamma=2$. The 
intrinsic column density was taken to be $1\times 10^{23}$~cm$^{-2}$ 
or the value from column three of Table~3 (whichever is 
larger).\footnote{In practice, {\sc pimms} is only able to work with 
column densities at $z=0$. Therefore, to include intrinsic column 
densities at $z>0$ in {\sc pimms} we 
used {\sc xspec} (Arnaud 1996) to find the column density at 
$z=0$ that produces equivalent absorption. Galactic absorption was
also taken into account in this process.}
Note that cosmological redshifting of the X-ray spectrum
can in some cases greatly reduce the effect of the intrinsic
absorption. This allows us to set more sensitive upper limits
on the 2--10~keV fluxes for some high redshift objects due
to the energy dependence of the \asca\ spectral response. 

\subsection{Column Density Constraints}

Since X-ray spectra cannot be modeled for most of the BALQSOs in our
sample, we have followed the method adopted by GM96 to place lower 
limits on their intrinsic column densities. Briefly, we assumed the 
underlying optical-to-X-ray continuum shape of a typical RQ QSO and 
added an intrinsic absorbing column until 
the predicted count rate matched our observed count 
rate or upper limit. This type of analysis relies upon the plausible but 
unproven assumption that our targets have typical RQ QSO X-ray continua 
viewed through an intrinsic absorbing column of gas. Comparisons of BALQSO and 
non-BALQSO optical/ultraviolet emission lines and continua show that these
two types of QSOs are remarkably similar, consistent with the view that these 
are not two inherently different classes of objects (e.g. Weymann et~al. 1991). 
We also take the gas in the absorbing column to have solar abundances 
(Anders \& Grevesse 1989), and this appears to be consistent with the
currently available data (see Arav 1997 and Hamann 1998). Further justification
for the assumptions that underlie this general method may be found in GM96. 
One additional issue not discussed by GM96 is that resonant absorption lines
may be a significant source of X-ray opacity due to the
large velocity dispersions of BALQSO outflows (compare with \S3 of 
Kriss et~al. 1996). We have neglected this effect for consistency with
all previous work and because there is presently no proof that the
BAL gas also causes the inferred X-ray absorption. A 
detailed treatment of resonant absorption line opacity in 
the X-ray band will be complex and dependent upon the details of the velocity 
dispersion in the BAL outflow, but this 
opacity is expected to modify the inferred column 
density by less than a factor of $\approx 2$ (J. Krolik, private
communication). Theoretical calculations examining the importance of 
this effect would be a valuable addition to the literature. 

The optical-to-X-ray spectral index, $\alpha_{\rm ox}$, is the slope of a 
nominal power law connecting the rest-frame flux density at 2500~\AA\ 
to that at 2~keV (QSOs with large values of $\alpha_{\rm ox}$ are those
that are X-ray faint). Typical RQ QSOs are observed to have 
$\alpha_{\rm ox}=1.51\pm 0.10$ (this value is from \S4.3 of 
Laor et~al. 1997 for an optical flux density at 2500~\AA; 
A. Laor, private communication).
The 2~keV flux density can be predicted given 
the rest-frame 2500~\AA\ flux density and an expected 
value for $\alpha_{\rm ox}$. Flux densities at 2500~\AA\ for PG~0043+039 and 
PG~1700+518 were obtained by interpolating the continuum flux density data 
of Neugebauer et~al. (1987) with a power law. For 
our non-PG BALQSOs, we derived the flux density at 
2500~\AA\ from the observed $B$ magnitude (see Table~1) using the flux 
zero point of Marshall et~al. (1983) and extrapolating along an assumed 
optical continuum power law with slope $\alpha_{\rm o}=0.5$. The Galactic 
extinction corrections used by GM96 and a $K$-correction for the 
power law were also included. 

An $\alpha_{\rm ox}$ value of 1.6 was used to predict an underlying 
rest-frame flux density at 2~keV for each BALQSO in our sample. 
We chose this value of $\alpha_{\rm ox}$ for consistency with 
GM96 and note that it is reasonably conservative (also see
the discussion in \S4.1 of GM96). We assumed that the 
underlying X-ray continuum shape was a power law with photon index 
$\Gamma$, and we then calculated the expected power-law normalization 
at 1~keV in the observed frame. We entered this model into {\sc xspec} 
(Arnaud 1996) allowing for the presence of absorption by a column 
of neutral gas intrinsic to the BALQSO. Galactic absorption was also
included with the column densities given in Table~1. Using the simulation 
routine {\sc fakeit} with the spectral response matrices for the \asca\ 
SIS0 and GIS3 detectors [described in \S10.4.1 of the 
AO-7 \asca\ Technical Description (AN 98-OSS-03)], a predicted count 
rate was generated. The intrinsic column density was then increased until 
the predicted count rate from the simulation no longer exceeded the measured 
count rate or upper limit. In this manner, an {\it upper\/} limit on the 
count rate yielded a {\it lower\/} limit on the intrinsic column density. 
We have focused on the SIS0 and GIS3 detectors because for
the standard pointing position our targets lie closer to the 
optical axes of these detectors. Combining like detectors
(SIS0/SIS1 or GIS2/GIS3) would not substantially improve the
column density constraints due to the fact that the vignetting 
for \asca\ is highly dependent upon off-axis angle and is thus
noticeably worse for SIS1 and GIS2. 
The results from this analysis are presented in Table~3.  
Since X-ray spectral shapes vary among the RQ QSO population, we 
have performed these calculations with both $\Gamma=1.7$ and 
$\Gamma=2.0$ in order to cover the typical range of photon index 
values (see Reeves et~al. 1997). Using PG~0043+039 as 
an example, we have also illustrated the dependence 
of the inferred column density upon the intrinsic 
$\alpha_{\rm ox}$ value in Figure~1.

Recent evidence indicates that some, if not all, BALQSOs have 
significantly attenuated optical/ultraviolet continua compared 
to non-BAL RQ QSOs (e.g. Goodrich 1997; see this paper for the
precise definition of `attenuation'). In this case, an underlying 
flux density at 2~keV that is predicted from the attenuated flux density 
at 2500~\AA\ will be artificially low. While this issue is difficult to 
address with precision at present, correcting for it would 
tend to strengthen our above results. Even larger intrinsic column
densities would be required to suppress the larger predicted X-ray fluxes. 
For two of our BALQSOs with near-infrared continuum flux densities in the 
literature, PG~0043+039 and PG~1700+518, we have examined this matter
by attempting to predict the underlying 2~keV flux density from the
1.69~$\mu$m flux density (again using the data of Neugebauer et~al. 1987). 
Laor et~al. (1994) argue that the 1.69~$\mu$m flux density is a reasonably 
good predictor of the 0.3~keV flux density, and the Laor et~al. (1997) 
data show that it is also a reasonably 
good predictor of the 2~keV flux density. We 
have computed $\alpha_{\rm ix}$, the spectral slope of a nominal power law
between 1.69~$\mu$m and 2~keV, for the 20 QSOs from Laor et~al. (1997) 
without intrinsic absorption. We find that $\alpha_{\rm ix}=1.29\pm 0.09$,
and that $\alpha_{\rm ix}$ has a comparable dispersion to $\alpha_{\rm ox}$
for the same set of QSOs. We have repeated our column density estimates
for PG~0043+039 and PG~1700+518 using $\alpha_{\rm ix}$ in place of 
$\alpha_{\rm ox}$, and we find that the inferred column density lower
limits are within a factor of two of those presented in Table~3. We
show our results for PG~0043+039 in Figure~2, and for PG~1700+518
we find an intrinsic column density $>8.5\times 10^{23}$~cm$^{-2}$
when $\alpha_{\rm ix}=1.29$ and $\Gamma=2.0$. 


\section{Notes on Individual Observations}

{\bf PG~0043+039:\/}
The optical and ultraviolet properties of this
BALQSO were recently studied in detail by Turnshek et~al. (1994), 
and the \asca\ analysis for it was straightforward. Our large inferred
X-ray column density combined with the intrinsic color excess of 
$E(B-V)\approx 0.11$ may suggest that the absorber is dust poor
(compare with \S3.1.1 of Turnshek et~al. 1994). 

{\bf 0043+008 (UM~275):\/}
Zamorani et~al. (1981) claimed that this BALQSO was detected in a
1.6~ks \einstein\ observation, although our analysis of the 
\einstein\ data made us skeptical of this claim. 
Wilkes et~al. (1994) also give only an \einstein\ upper limit. 
0043+008 is undetected in our much deeper (23~ks) \asca\ observation.  
Based on our analysis, it appears that the claimed \einstein\ 
detection was actually an unrelated source lying $2.1^{\prime}$ 
from the precise optical position. In order 
to obtain the tightest possible constraints on 
the SIS and GIS count rates for 0043+008, we have excluded this source
from the target regions before calculating the count rate upper 
limits. This source does not cause serious confusion, but there
was a statistical photon excess in the target region for
GIS3 that was not consistent with a point source or
the optical position of 0043+008. 
We suspect this excess is due to imperfect removal 
of all the photons from the \einstein\ source. 

{\bf 0226--104:\/}
Despite the fact that this BALQSO has been intensively studied
(e.g. Korista et~al. 1992), the coordinates stated for it in 
the literature are inconsistent with each other. 
We have used the coordinates
$\alpha_{2000}=02^{\rm h} 28^{\rm m} 39^{\rm s}.2$,
$\delta_{2000}=-10^\circ 11^\prime 10.0^{\prime\prime}$ 
(T. Barlow, private communication). 
0226--104 was detected in all but the SIS1 detector as a point 
source (the SIS1 non-detection can be understood as due to the
larger off-axis angle of 0226--104 in this detector). 
We excluded counts from two nearby sources before calculating 
the count rates. 

{\bf IRAS~07598+6508:\/}
This BALQSO was marginally detected by \rosat\ with a count rate of
$\approx 1.8\times 10^{-3}$~count~s$^{-1}$, but it was impossible to 
determine whether the observed X-ray emission was associated with 
accretion activity or a circumnuclear starburst (Lawrence et~al. 1997).
There was an X-ray source coincident with the optical position
of IRAS~07598+6508 in the GIS3 detector, and the target region 
counts were $8.6\sigma$ above background for the
full band image (the source was also seen in the hard band image). 
There were photon excesses in the target regions for SIS0 and
SIS1, but the lack of distinct point sources at the optical
position led us to conservatively treat these as upper limits. 
The sole detection in the GIS3 detector might be understood 
as due to the better high-energy response of the GIS detectors 
and the smaller off-axis angle of GIS3 compared to GIS2.  

Our lower limit on the absorption column density is $\approx 20$
times higher than that of GM96, and this suggests that the
dust-to-gas ratio in the absorbing material may be extremely
small (compare with \S5.2 of GM96).

{\bf Mrk~231:\/}
This object has the lowest redshift in our sample and has 
sometimes been classified as a Seyfert~1, but we feel its
large luminosity ($L_{\rm Bol}>10^{46}$~erg~s$^{-1}$) and 
strong absorption lines justify its inclusion 
(e.g. it is also included in the sample of Boroson \& Meyers 1992). 
Mrk~231 has recently been studied with \asca\ by Iwasawa (1999)  
and Turner (1999). Our spectral analysis is consistent with the analyses 
in these papers and thus we do not detail it here. In Table~3 we 
give the column density discussed in \S4.1.2 of Iwasawa (1999).
However, even the fitted \asca\ column density may be 
substantially smaller than the true column density to the
black hole region due to possible electron scattering and partial
covering effects in this extremely complicated system
(see \S4 for further discussion).

{\bf PG~1700+518:\/}
There were no \asca\ sources close to the optical position for this 
BALQSO, and we were able to place tight upper limits on the count rate 
for all but the SIS1 detector. A statistical photon excess in the SIS1 
target region was not from any obvious point source and thus we do not 
consider it a detection.  
A 40~ks \rxte\ observation of PG~1700+518 has also been performed 
with principal investigator P. Green. Given that our \asca\ 
observation is $\simgt 15$ times more sensitive than the \rxte\ 
observation, we expect an \rxte\ non-detection.  

{\bf LBQS~2111--4335:\/}
There was clearly no point source at the optical position of the 
BALQSO. However, a ninth magnitude A star (HD202042) was coincident
with an X-ray source detected 2.2$^{\prime}$ from LBQS~2111--4335
(the A star may have a late-type binary companion that creates 
most of the X-ray emission). We excluded 
the counts from this source before calculating the count rate
upper limits. Even after excluding the source from the extraction 
region, photon counts above the noise level were still evident 
in the SIS0 detector. However, we do not consider this to be
a detection.

{\bf PHL~5200:\/}
The \asca\ data for this object were first presented by Mathur et~al. (1995). 
This BALQSO was clearly detected in all but the GIS2 detector. We find 
that the results of spectral analysis are highly sensitive to the details
of the background subtraction. Given this sensitivity, our analyses suggest 
that any column density from 0--$5\times 10^{23}$~cm$^{-2}$ is consistent 
with the ASCA data (even at 68\% confidence). Thus, while the \asca\ data 
certainly do not rule out the presence of a large column density absorber, 
they do not convincingly show one to be present either.
 

\section{Conclusions and Future Prospects}

We have presented the first moderate-sized sample of sensitive BALQSO 
observations in the 2--10~keV band. This band is potentially an 
important one for BALQSO spectroscopy due to the ability of $>2$~keV 
X-rays to penetrate large column densities, and the properties of our 
sample are likely to be more representative of the behavior of the
BALQSO population as a whole than those derived from the two 
single-object \asca\ studies to date (see \S1). While we 
detect a significantly larger fraction of our objects than
GM96, we find that in general even the optically brightest 
BALQSOs known are extremely weak in the 2--10~keV band. 
Assuming that our BALQSOs have typical RQ QSO X-ray continua and are 
weak due to intrinsic X-ray absorption, we find column densities
$\simgt 5\times 10^{23}$~cm$^{-2}$ in several cases. These are the
largest X-ray column densities yet inferred for BALQSOs, being about
an order of magnitude larger than the lower limits of GM96. 
If the same outflowing gas makes both the X-ray and ultraviolet 
absorption, our improved column density limits also raise the implied  
mass outflow rate and kinetic luminosity by about an order of magnitude. 

Alternatively, it is possible that BALQSOs are intrinsically 
X-ray weak, perhaps due to an orientation dependence of the X-ray 
continuum flux. Krolik \& Voit (1998) have discussed the 
possibility of optical/ultraviolet continuum anisotropy in 
BALQSOs (due to accretion disk limb-darkening which dims objects
viewed equatorially), but they suggest that anisotropy of 
the X-ray emission is likely to be weaker. If the power-law 
X-ray emission of RQ QSOs indeed originates in an optically 
thin accretion disk corona near a Kerr black hole, the intrinsic 
equatorial emission may in fact be enhanced by relativistic 
effects (e.g. Cunningham 1975). 

An important corollary of our work is that optical brightness
(in the $B$ or $R$ bands) is {\it not\/} a good predictor of X-ray
brightness for BALQSOs. Two of the four BALQSO X-ray detections in our 
sample (0226--104 and PHL~5200) are actually among our optically faintest 
objects. It appears that the prototype of the class, PHL~5200 (Lynds 1967),
has an unusually high X-ray flux for a BALQSO given its optical flux, and thus
its X-ray properties cannot be taken as representative of the 
BALQSO population.
In fact, its X-ray brightness is typical of non-BALQSOs 
of its optical magnitude, and we do not find the evidence
for a large intrinsic column density in this object to be 
compelling (see \S3). 
Finally, all four of our detected BALQSOs have notably 
high optical continuum polarizations (2.3--5\%; 
Schmidt, Hines \& Smith 1997; 7 of our 8 BALQSOs have 
optical continuum polarization data in the literature), and 
it is worth investigating if high-polarization BALQSOs tend to 
be the X-ray-brighter members of the class. 
We refrain from making a formal
claim about this issue at present due to our limited sample 
size, but another example of this phenomenon may
be IRAS~13349+2438 (see \S3.1.3 of Brandt et~al. 1997). 
High optical continuum polarization and X-ray flux could occur 
together if the direct lines of sight into the X-ray nuclei
of BALQSOs were usually blocked by extremely thick material
(say $\sim 10^{25}$~cm$^{-2}$). 
In this case, one could only hope to detect X-rays when there 
is a substantial amount of electron scattering in the nuclear 
environment by a `mirror' of moderate Thomson depth.
Then, any measured X-ray column density for a highly polarized BALQSO
(e.g. Mrk~231 and PHL~5200) might not reflect the true
column density to the black hole region but rather the column 
density along the electron-scattered path (compare with \S5
of Goodrich 1997). We have 
also looked for other common properties of our four 
X-ray detected BALQSOs and none is apparent. For example, they 
are not all `PHL 5200-like' BALQSOs (see Turnshek et~al. 1988).

Aside from the physical constraints placed upon the X-ray column 
densities in BALQSOs, our results also have important practical 
implications for future X-ray spectroscopy of these objects. 
Consider the representative case of PG~1700+518. The \asca\ SIS and GIS count 
rates for this BALQSO are $<3.2\times 10^{-3}$~count~s$^{-1}$, and its 2--10~keV 
observed flux is constrained to be $<3\times 10^{-13}$~erg~cm$^{-2}$~s$^{-1}$. 
Using {\sc pimms}, this translates into an observed \axaf\ ACIS-I count 
rate of $<9\times 10^{-3}$~count~s$^{-1}$. While this BALQSO may be detectable 
with \axaf, high-quality spectroscopy of it will be difficult and perhaps 
impossible (assuming no strong variability). Even in the most optimistic 
case, where the flux is just below our upper limit, it will take 
$\approx 110$~ks to obtain the $\approx 1000$ counts needed for
moderate-quality spectroscopy.\footnote{The $1000$ count and 10\,000 
count criteria are those given in \S1.5 of the {\it AXAF Proposers' Guide\/}, 
and we find that these criteria work well in practice. Similar criteria are 
adopted in figures~29--33 of the {\it XMM Users' Handbook\/}.} 
A high-quality ($\approx 10\,000$ count) spectrum would require an 
immodest $\simgt 10^6$~s of \axaf\ time. For the \xmm\ EPIC-PN, the 
count rate will be $<2.6\times 10^{-2}$~count~s$^{-1}$. 
Spectroscopy may be more tractable in this case, but it will still prove
inordinately expensive if the flux is a factor of a few times smaller
than our \asca\ upper limit. 
We also comment that some of our implied column density lower limits are 
close to becoming optically thick to Thomson scattering. If the X-ray 
column densities in many BALQSOs are optically thick to Thomson scattering, 
this will make it difficult to improve the above situation by observing at
even higher energies. 


\section{Acknowledgments}

We thank 
M. Elvis, 
E. Feigelson, 
E. Gotthelf, 
J. Krolik,
A. Laor, 
K. Mukai, 
D. Schneider and
an anonymous referee
for helpful discussions. 
This paper is based upon work supported by 
NASA grant NAG5-4826 (SCG),
the NASA LTSA program (WNB),  
NASA contract NAS-38252 (RMS), and
NASA grants NAG5-3249 and NAG5-3390 (SM). 


\pagebreak

\pagebreak


\begin{deluxetable}{lcccccc}
\tablenum{1}
\tablewidth{0pt}
\tablecaption {BALQSO General Information and Observing Log}
\small
\tablehead{ 
\colhead{Target}            & 
\colhead{}                  & 
\colhead{}                  &
\colhead{Galactic $N_{\rm H}$}   &
\colhead{Obs. }      & 
\colhead{SIS Exp. (ks)/} &
\colhead{GIS}\\
\colhead{Name$^{\rm a}$}                & 
\colhead{$z$}                           & 
\colhead{$B/R^{\rm b}$}                         & 
\colhead{/($10^{20}$~cm$^{-2}$)}        &
\colhead{Date}                          & 
\colhead{CCD Mode$^{\rm c}$}            &
\colhead{Exp. (ks)$^{\rm d}$}
}
\startdata
PG~0043+039  & 
0.384      &
15.9$^{\rm (1)}$/15.6$^{\rm (2)}$      &
3.0$^{\rm (9)}$      & 
21 Dec 96      & 
22.7/2 CCD &
28.3     \nl
0043+008  & 
2.146     &
18.4$^{\rm (3)}$/17.7$^{\rm (4)}$      &
2.0$^{\rm (10)}$      & 
09 Jul 98      & 
23.0/1 CCD  &
23.9      \nl
0226--104  & 
2.256      &
17.3$^{\rm (5)}$/15.2$^{\rm (4)}$     &
2.6$^{\rm (11)}$      & 
17 Jul 98      & 
24.4/1 CCD  &
23.0     \nl
IRAS~07598+6508 (L) & 
0.149      &
14.5$^{\rm (6)}$/13.1$^{\rm (3)}$     &
4.5$^{\rm (12)}$    & 
29 Oct 96      & 
37.2/1 CCD  &
41.7      \nl
Mrk~231 (L) &  
0.042      &
14.5$^{\rm (7)}$/13.2$^{\rm (7)}$     &
1.0$^{\rm (10)}$      & 
05 Dec 94      & 
15.7/2 CCD  &
20.0    \nl
PG~1700+518 (L) & 
0.292      &
15.4$^{\rm (1)}$/14.4$^{\rm (3)}$    &
2.5$^{\rm (9)}$     & 
24 Mar 98      & 
19.2/1 CCD  &
20.6      \nl
LBQS~2111--4335  & 
1.708      &
16.7$^{\rm (8)}$/16.0$^{\rm (3)}$     &
4.2$^{\rm (13)}$      & 
23 Apr 98      & 
21.5/1 CCD  &
22.6      \nl
PHL~5200& 
1.981   &
18.0$^{\rm (5)}$/17.4$^{\rm (4)}$     &
4.8$^{\rm (11)}$      & 
21 Jun 94      & 
10.2/2 CCD  &
16.4      \nl
\enddata
\tablenotetext{a}{An `L' in parentheses after the target
 name indicates that this BALQSO is 
known to show low-ionization BALs (see Boroson \& Meyers 1992 and references therein). 
The optical spectrum for LBQS~2111--4335 presented in Morris et~al. (1991) suggests
that broad Al~{\sc iii} absorption may be present. However, the current data do not
allow a definitive classification (S. Morris, private communication).}
\tablenotetext{b}{Magnitudes have not been corrected for BALs. 
The $R$ magnitude is defined as noted in the table references.}
\tablenotetext{c}{All SIS observations were done in Faint mode.}
\tablenotetext{d}{All GIS observations were done in Pulse Height (PH) mode.} 
\tablenotetext{}{$^1$Schmidt \& Green (1983)
$^2$Neugebauer et~al. (1991); $R$ computed from continuum fluxes 
$^3$McMahon \& Irwin (1992)---APM catalogue; photographic $E$
$^4$Sirola et~al. (1998); Gunn r 
$^5$Stocke et~al. (1992)
$^6$Turnshek et~al. (1997)
$^7$Hutchings \& Neff (1997); Johnson $R$ 
$^8$Hewett, Foltz \& Chaffee (1995)}
\tablenotetext{}{$^9$Lockman \& Savage (1995)
$^{10}$Elvis, Lockman \& Wilkes (1989)
$^{11}$Stark et~al. (1992)
$^{12}$Murphy et~al. (1996)
$^{13}$Heiles \& Cleary (1979)}
\end{deluxetable}


\begin{deluxetable}{lccccc}
\tablenum{2}
\tablewidth{0pt}
\tablecaption {\asca\ Detector Count Rates$^{\rm a}$/($10^{-3}$~photons s$^{-1}$). }
\small
\tablehead{ 
\colhead{Target Name}            & 
\colhead{SIS0} &
\colhead{SIS1} & 
\colhead{GIS2} &
\colhead{GIS3} &
\colhead{Observed $F_{2-10}^{\rm b}$}
%
}
\startdata
PG~0043+039  & 
$<1.6$      &
$<1.6$      &
$<1.8$      & 
$<1.9$      &
$<2$ \nl
0043+008  & 
$<2.2$      &
$<2.1$      &
$<2.3$      & 
$<4.4^{\rm c}$       &
$<1$ \nl
0226--104  & 
2.5      &
$<2.2$     &
2.2      & 
5.1      &
$1$ \nl
IRAS~07598+6508  & 
$<2.3^{\rm c}$      &
$<3.3^{\rm c}$     &
$<1.7$      & 
$4.2$   &
$\simlt 4$    \nl
Mrk~231 &  
7.2      &
6.4      &
6.0      & 
7.1      &
8 \nl
PG~1700+518  & 
$<1.9$      &
$<3.2^{\rm c}$    &
$<2.3$      & 
$<2.4$      & 
$<3$    \nl
LBQS~2111--4335  & 
$<4.0^{\rm c}$      &
$<2.1$      &
$<2.3$      & 
$<2.5$      &
$<2$ \nl
PHL~5200 & 
3.8      &
6.3      &
$<2.9$   & 
6.3      & 
$1$ \nl
\enddata
\tablenotetext{a}{SIS/GIS count rates are for 0.6--9.5~keV/0.9--9.5~keV, 
and count rate upper limits are for $3\sigma$ unless otherwise noted.} 
\tablenotetext{b}{$F_{2-10}$ is the observed 2--10~keV flux 
in units of $10^{-13}$~ergs~cm$^{-2}$~s$^{-1}$
computed using a power law with $\Gamma=2$. For the intrinsic
column density we use either $1\times 10^{23}$~cm$^{-2}$ or
the value from column three of Table~3 (whichever is larger). The
Galactic column density is also included (see \S2.2).} 
\tablenotetext{c}{Photon excess above $3\sigma$ in the target region, 
but no evidence of a point source at the optical position of the 
target. See \S3 for discussion.} 
\end{deluxetable}


\begin{deluxetable}{lccc}
\tablenum{3}
\tablewidth{0pt}
\tablecaption {Limits on Intrinsic Absorption$^{\rm a}$}
\small
\tablehead{ 
\colhead{Target}                            & 
\colhead{Assumed X-ray}                     & 
\colhead{SIS0/GIS3 Intrinsic}               \\
\colhead{Name}                                   & 
\colhead{Photon Index}                           & 
\colhead{$N_{\rm H}$/($10^{22}$~cm$^{-2}$)$^{\rm b}$}      & 
}
\startdata
PG~0043+039  & 
1.7     & 
$>47$/$>40$	&
	\nl

	&
2.0     & 
$>33$/$>26$	&
       \nl
0043+008  & 
1.7     & 
$>1.8$/$\cdots$	&
	\nl

	&
2.0     & 
$\cdots$/$\cdots$	&
       \nl
0226--104  & 
1.7     & 
$\cdots$/$\cdots$	&
	\nl

	&
2.0   & 
$\cdots$/$\cdots$    &
       \nl
IRAS~07598+6508  & 
1.7     & 
$>72$/$50$  &
	\nl

	&
2.0   & 
$>56$/$36$	&
  \nl
Mrk~231$^{\rm c}$ &  
$1.8$    & 
$4$	&
	\nl
PG~1700+518  & 
1.7     & 
$>59$/$>53$	&
	\nl

	&
2.0   & 
$>44$/$>36$	&
      \nl
LBQS~2111--4335  & 
1.7     & 
$>2.6$/$>3.9$	&
	\nl

	&
2.0   & 
$\cdots$/$\cdots$	&
    \nl
PHL~5200 & 
1.7     & 
$\cdots$/$\cdots$	&
	\nl

	&
2.0   & 
$\cdots$/$\cdots$	&
       \nl
\enddata
\tablenotetext{a}{For $\alpha_{\rm ox}=1.6$, unless otherwise noted.} 
\tablenotetext{b}{An ellipsis indicates that the intrinsic column density was consistent with 0~cm$^{-2}$.} 
\tablenotetext{c}{For details on the photon index and column density
constraints for Mrk~231 see \S3, Iwasawa (1999), and Turner (1999).} 
\end{deluxetable}


\begin{figure}
\epsscale{0.5}
\plotfiddle{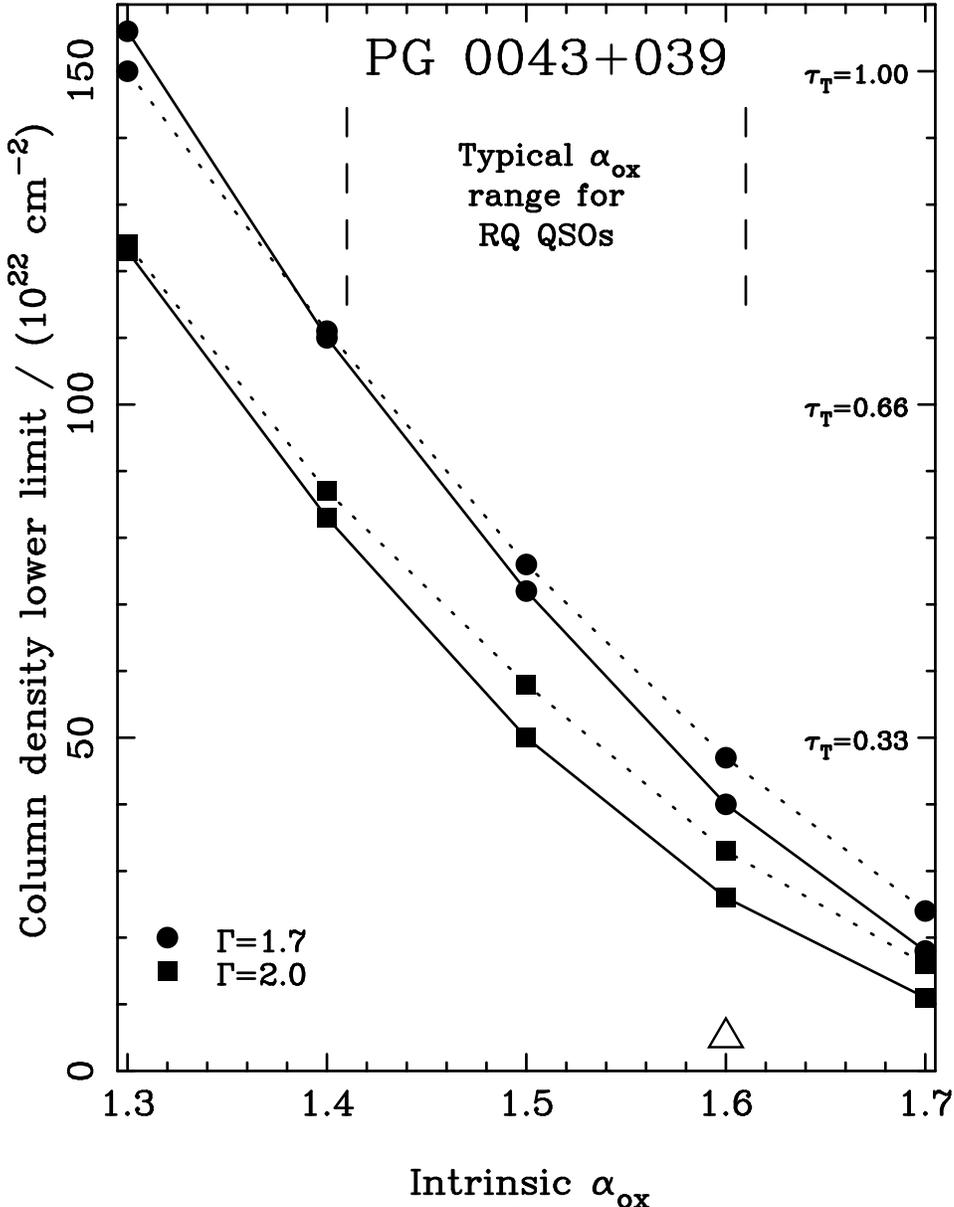}{340pt}{0}{70}{70}{-210}{-120}
\vspace{0.9in}
\caption{Column density lower limits for PG~0043+039 derived following the method 
described in the text. We show the inferred column density lower limit 
as a function of the intrinsic (i.e. corrected for X-ray absorption) $\alpha_{\rm ox}$. 
The square data points are for an X-ray photon index of $\Gamma=2.0$ and the circular 
dots are for an X-ray photon index of $\Gamma=1.7$. The dotted lines show the constraints 
from SIS0, and the solid lines show the constraints from GIS3. 
The open triangle at an intrinsic $\alpha_{\rm ox}=1.6$ illustrates the typical 
BALQSO column density lower limit found by GM96 based on \rosat\ data (although 
GM96 did not present \rosat\ data for PG~0043+039). The two vertical dashed
lines show the typical range of $\alpha_{\rm ox}$ for RQ QSOs (see the text). The 
numbers along the right hand side of the plot show the Thomson optical depth of the
corresponding column density. Note that our inferred column densities are within 
a factor of $\approx 3$ of being optically thick to Thomson scattering. 
\label{fig1}}
\end{figure}

 

\begin{figure}
\epsscale{0.5}
\plotfiddle{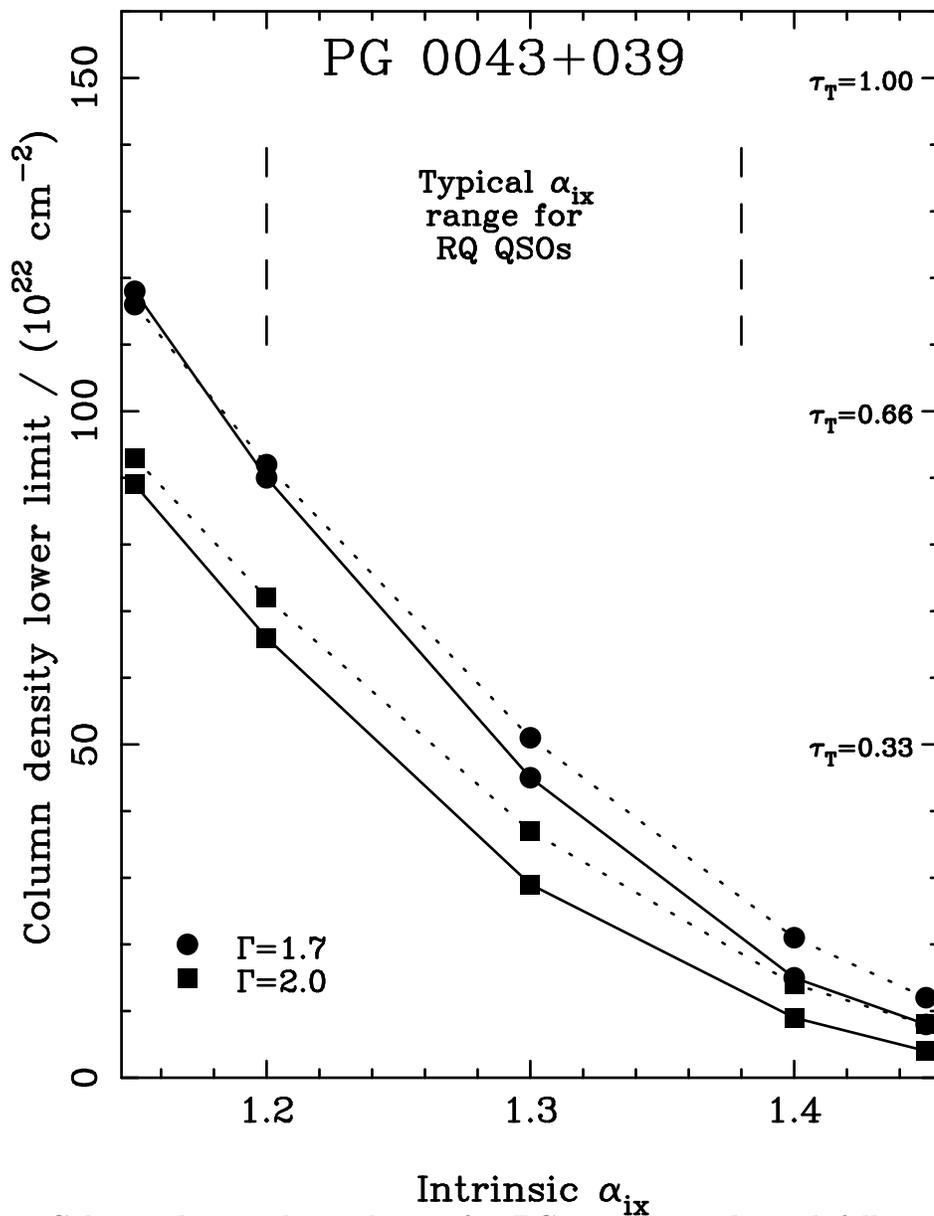}{340pt}{0}{70}{70}{-210}{-120}
\vspace{0.9in}
\caption{Column density lower limits for PG~0043+039 derived following the method 
described in the text. We show the inferred column density lower limit 
as a function of the intrinsic (i.e. corrected for X-ray absorption) $\alpha_{\rm ix}$.
The symbols and line styles are as for Figure~1, and the typical range of 
$\alpha_{\rm ix}$ is discussed in the text. 
\label{fig1}}
\end{figure}

 

\end{document}